\documentstyle[preprint,aps,epsfig]{revtex}
\advance\hsize by 0.0truecm  \hoffset=0.7cm
\begin{document}
\bibliographystyle{prsty}
\newcommand{\beq}{\begin{equation}}
\newcommand{\eeq}{\end{equation}}
\newcommand{\bea}{\begin{eqnarray}}
\newcommand{\eea}{\end{eqnarray}}
\newcommand{\ms}{m_{\rm s}}
\newcommand{\alf}{{\bar a}}
\newcommand{\bet}{{\bar b}}
\newcommand{\dmuu}{{\partial^\mu}}
\newcommand{\dmud}{{\partial_\mu}}
\newcommand{\lb}{\hfil\break }
\newcommand{\qeq}[1]{eq.\ (\ref{#1})  }
\newcommand{\qeqs}[2]{eqs.\ (\ref{#1}) and (\ref{#2}) }
\newcommand{\queq}[1]{(\ref{#1})}
\newcommand{\qutab}[1]{Tab. (\ref{#1}) }
\newcommand{\qufig}[1]{Fig. (\ref{#1}) }
\newcommand{\qtab}[1]{Tab. (\ref{#1})}
\newcommand{\dsl}{ \rlap{/}{\partial} }
\newcommand{\vsl}{\rlap{V}{\hskip1pt/}}
\newcommand{\asl}{\rlap{A}{\hskip2pt/}}
\newcommand{\ssl}{\rlap{s}{/} }
\newcommand{\xsl}{\rlap{x}{/} } 
\newcommand{\ps}{ \rlap{/}{p} }
\newcommand{\half}{\frac{1}{2}}
\newcommand{\quref}[1]{\cite{ins:#1}}
\newcommand{\qref}[1]{Ref.\ \cite{bolo:#1}}
\newcommand{\wu}{\sqrt{3}}
\newcommand{\nn}{\nonumber \\ }
\newcommand{\Y}{\ {\cal Y}}
\newcommand{\Sp}{{\rm Sp\ } }
\newcommand{\Tr}{{\rm Tr}_{\gamma\lambda_c}}
\newcommand{\Trto}{{\rm Tr}_{\gamma\lambda_c(to)}}
\newcommand{\tr}{{\rm tr} }
\newcommand{\sign}{{\rm sign} }
\newcommand{\Spto}{{\rm Sp_{(to)}\ } }
\newcommand{\linie}{\ \vrule height 14pt depth 7pt \ }
\newcommand{\intT}{\int_{-T/2}^{T/2} }
\newcommand{\Nc}{N_{\rm c}}
\newcommand{\Ne}{$N_{\rm c}$\ }
\newcommand{\gs}{$g_{\rm A}^{0}$}
\newcommand{\gt}{$g_{\rm A}^{3}$}
\newcommand{\go}{$g_{\rm A}^{8}$}
\newcommand{\Gs}{g_{\rm A}^{0}}
\newcommand{\Gt}{g_{\rm A}^{3}}
\newcommand{\Go}{g_{\rm A}^{8}}
\newcommand{\MeV}{{\rm MeV}}
\newcommand{\lag}{Lagrangian }
\newcommand{\uas}{$U_A(1)$ symmetry }
\newcommand{\atan}{\rm arctan }
\newcommand{\intq}{\int {d^4q\over {(2\pi)}^4}  }
\newcommand{\daa}{D_{88}^{(8)}(A) }
\newcommand{\dari}{\sum_{i=1}^3 D_{8i}^{(8)}(A) R_i }
\newcommand{\dara}{\sum_{a=4}^7 D_{8a}^{(8)}(A) R_a }
\newcommand{\daiai}{\sum_{a=1}^3 D_{8i}^{(8)}(A) D_{8i}^{(8)}(A) }
\newcommand{\daaaa}{\sum_{a=4}^7 D_{8a}^{(8)}(A) D_{8a}^{(8)}(A) }
\newcommand{\bit}{\begin{itemize}}
\newcommand{\eit}{\end{itemize}}
\newcommand{\tot}{ {3\over 2} }
\newcommand{\ba}{\begin{array} }
\newcommand{\ea}{\end{array} }
\newcommand{\ksl}{ \rlap{/}{k} }
\newcommand{\pslash}{ \rlap{/}{p} }
\newcommand{\ddmu}{\partial_\mu }
\newcommand{\dumu}{\partial^\mu }
\newcommand{\vx}{ {\vec x}  }
\newcommand{\seff}{S_{\rm eff} }
\newcommand{\phans}{\phantom{abc} }
\newcommand{\phan}{\phantom{abcdefgh} }
\newcommand{\A}{ {\cal R} }
\newcommand{\njlm}{Nambu--Jona-Lasinio model\ }
\newcommand{\njl}{Nambu--Jona-Lasinio\ }
\newcommand{\oon}{$1/N_c$}
\newcommand{\GeV}{{\rm GeV}}
\newcommand{\disp}{\displaystyle }
\newcommand{\vy}{ {\vec y} }
\newcommand{\vz}{ {\vec z} }
\newcommand{\dtau}{\partial_\tau}
\newcommand{\Mhat}{{\hat M}}
\newcommand{\cor}{{\cal C}_B(T)}
\newcommand{\brax}{\langle \vx,t_x\mid}
\newcommand{\bracx}{\langle x\mid}
\newcommand{\ketsx}{\mid x\rangle}
\newcommand{\bray}{\langle \vy,t_y\mid}
\newcommand{\braz}{\langle \vz,t_z\mid}
\newcommand{\ketx}{\mid\vx,t_x\rangle}
\newcommand{\kety}{\mid\vy,t_y\rangle}
\newcommand{\ketz}{\mid\vz,t_z\rangle}
\newcommand{\defined}{\stackrel{\rm def}{=}}
\newcommand{\signum}{{\rm sign} }
\newcommand{\ket}[1]{\mid{#1}\rangle}
\newcommand{\bra}[1]{\langle{#1}\mid}
\newcommand{\mata}{\left( \ba{c} }
\newcommand{\mate}{\ea \right) }
\newcommand{\diag}{{\rm diag}}
\newcommand{\sumi}{ \sum_{i=1}^3 }
\newcommand{\suma}{ \sum_{a=4}^7 }
\renewcommand{\arraystretch}{1.5}
\newcommand{\fm}{{\rm fm}}
\newcommand{\ohatz}{{\hat O}_\mu(z)}
\newcommand{\square}{\partial_i^2}
\newcommand{\PP}{ {\rm PP} }
\newcommand{\moment}{ \Theta_U }
\newcommand{\bmu}{{\bar\mu}}
\newcommand{\bmo}{{\bar m}_0}
%
%
%
\preprint{SUNY-NTG-97-50  \\ 
\ \ \ \ \  LA-UR-97-4446  }
 
\title{
Instanton-induced charm contribution \\
 to polarized deep-inelastic scattering
}
\author{    Andree Blotz $^{(1)}$ and 
            Edward Shuryak $^{(2)}$ 
            }
\address{(1)
 Theoretical Division, 
             T5 MS B283, LANL, Los Alamos, NM 87545, USA \\ 
              email:blotz@t5.lanl.gov      }

 \address{(2) 
Department of Physics/Astronomy,
 State University of New York at Stony Brook, \\
NY  11794-3800, USA  \\    email:shuryak@dau.physics.sunysb.edu 
      }
\date{ \today  } 
\maketitle
\begin{abstract}
Recent data on B decays involving $\eta'$ may be explained if the 
singlet axial 
current of charmed quarks has a large matrix element to  $\eta'$, and
instantons were shown to be able to generate this effect. We study the 
magnitude of 
charm contributions to nucleon polarized structure functions
generated in a similar way. Comparing the charm contribution,
which is related to a dim(6) gluonic operator, to that of light quarks,
which are related by the anomaly equation to $G\tilde G$,  
we found
that $\Delta c/\Delta\Sigma$ = -(0.2-0.08).
Future experiments like COMPASS at CERN 
identifying charm production in DIS   
can measure 
this ``intrinsic polarized charm'' component of
the nucleon.

\end{abstract}
 \pagebreak

1.  It was believed that non-perturbative phenomena in QCD can generate 
light quark (and gluon)  ``sea'' in nucleon structure functions, while
the admixture of heavy flavors such as charm is (i) very small due to their
mass, and that (ii) it should 
 be produced by  perturbative effects alone.
Although the first part of the
argument (suppression by the power of charm mass
squared) is correct, it can be counterbalanced by  
sufficiently strong 
non perturbative gluon fields.
Small-size instantons present in the QCD vacuum have fields $G\sim 1 GeV^2$,
which is not much smaller than $m_c^2\sim 2 GeV^2$, and so in principle
one may expect a sizable ``intrinsic charm'' of at least some hadrons
(as anticipated by Brodsky et al. for a long time \cite{Brodsky:1980}). 

  Probably the first indication of that came from
 recent CLEO data on 
$B\rightarrow \eta' K$ and inclusive $B\rightarrow \eta'+...$
decays \cite{Behrens:1998dn}, which can be explained \cite{Halperin:1997a} 
if $\eta'$ has
such a charm component. Quantitative estimates of this effect 
 -- based on an instanton 
liquid model for the non perturbative vacuum -- were made in
\cite{shuzhi:1997}, which concluded that indeed current information about instantons
 lead to surprisingly large 
 charm content of the 
$\eta'$, comparable to 
what is needed to explain the CLEO data.

 It was further speculated in  \cite{shuzhi:1997} that other processes related to
 axial current of charm quarks may also show similar effects.
One of them is the charm contribution to
the polarized deep inelastic
scattering (DIS), which we discuss in this work.
Crude estimates given in  \cite{shuzhi:1997}
suggested that it may be even comparable to 
that of all light quarks together
(which is, we remind, already strongly suppressed, as compared to the naive unit
value
for the ``spin content'' of the nucleon).

In the present paper we address 
the question of an intrinsic polarized charm component 
directly, by evaluating  relevant three-point correlation functions
in the vacuum. We use the same 
instanton liquid models as
in ref. \cite{shuzhi:1997}. The results however
 are about an order of magnitude smaller
than the estimates presented there. Nevertheless, 
they are still very different from the perturbative predictions,
which imply that the charm sea is mostly unpolarized.

 2. Let us first recall few
 standard formulae  for polarized DIS on a nucleon. 
\footnote{See e.g.
\cite{Ramsey:1997iu} for a recent review and nomenclature.} 
The first moment of the  polarized structure function
\beq  
       \Gamma_1^p(Q^2) = \int_0^1 dx g_1(x,Q^2)  
 \label{dis1} 
\eeq 
can be related to the matrix element of the axial current 
by  the standard OPE treatment. The answer (including
  charm component $\Delta c$) can be written as \cite{Halperin:1997b}
 \beq    
  \Gamma_1^p(Q^2) = C_{NS}(Q^2) \left(  
{g_A^3 \over 12}  +  { g_A^8 \over 36} \right)  
 + {1 \over 9}  C_{S}(Q^2)  \left(   \Sigma   + 2  \Delta c
\right) 
\eeq 
where  $\Delta c$ is
\beq 
     s_\mu \Delta c(\mu^2) = \langle p,s \mid \left(  {\bar c} \gamma_\mu \gamma_5
                            c  \right)_{\mu^2}  
      \mid p,s \rangle      
\label{matc}
\eeq 
at a relevant mass scale $\mu^2$.  
Here the coefficients $ C_{NS}(Q^2), C_{S}(Q^2)$ account for perturbative
corrections of the form $1 - \alpha_s/\pi + \dots$ \cite{Ramsey:1997iu}.  
The measureable singlet axial coupling constant can be defined as
\beq  
       g_A^{(0)} = \Delta \Sigma  + 2 \Delta c  
\label{gaodef} 
\eeq 
where the first term is the contribution of all light quarks
 (u,d,s) together,
defined as $\Delta\Sigma=\Delta u + \Delta d + \Delta s$.

Both charm and light quark 
contributions can be calculated  from the 
divergence of the axial current: for each flavor one has (in Euclidean space) 
\beq    -\partial_\mu ( {\bar q}_E \gamma_\mu \gamma_5 q_E) 
        = 2 m_q  {\bar q}_E \gamma_5 q_E
    +  {1 \over 16 \pi^2 } 
        i  (G_{\mu\nu} {\tilde G_{\mu\nu}})_E 
\label{diveq}
\eeq 
However, the role of these two terms is different for each flavor.
For light quarks one can take the chiral limit $m\rightarrow 0$ and
ignore the first term in the r.h.s. The OPE expansion of the first term
in $1/m$ was done by  
Halperin and Zhitnitsky \cite{Halperin:1997a} and it shows that the anomaly
 term actually cancels,  
 while the leading 
term in the heavy quark limit is 
\beq  
        - \partial_\mu ( {\bar c}_E \gamma_\mu \gamma_5 c_E) 
          =   +i  { 1 \over 16 \pi^2 m_c^2 }  f^{abc}   
        G_{\mu\nu}^a {\tilde G}_{\nu\alpha}^b G_{\alpha\nu}^c 
                   + {\cal O}(G^4/m_c^4)  
\eeq    
Note that it is indeed suppressed by $1/m_c^2$, as it should.

3.Using lattice gauge configurations (or a model which defines gauge
fields in some way, as we do below), one can correlate these two 
 gluonic operators with the nucleon and thus evaluate the
matrix elements in question, $\Delta \Sigma$ and $ \Delta c$. 
Information about them is contained in 
  the three point correlation functions 
\beq     \Pi_{\mu\{\nu\}}^{\pm}(-x/2,x/2,y)   =
          \Gamma_{\beta\alpha}^{ \{\nu\} \pm}    \ \
         \langle 0 \mid {\cal T} \{  {\eta}_\alpha (-x/2)
         j_{\mu 5} (y) {\eta^\dagger}_\beta (x/2) \}  \mid 0 \rangle 
\label{corrfun}
\eeq 
where $\mu$ is the component of the axial current 
$ j_{\mu 5}(y)= {\bar q} \gamma_\mu \gamma_5 q$ and $\{\nu\}$ 
a multiple n-index for the projection operator  
$\Gamma^{ \{\nu\} \pm} = \gamma_{\nu 1}\dots\gamma_{\nu n} P_\pm $,
$P_\pm=P_R \pm P_L$. The ${\eta}_\alpha(x),{\eta^\dagger}(x)$ 
are the so called Ioffe currents, local
 operators with the quantum numbers of the nucleon 
\footnote{Since the latter do not contain charm quarks, the 
contribution of charm quarks is only via the disconnected, 
OZI violating, diagrams.} or in a simplified consideration the quarks.

  Before we go for the evaluation of the three-point functions, let us 
qualitatively explain the
$signs$ and magnitude of the gluonic operators for an
 instanton (anti-instanton) configuration. Introducing the topological
charge $Q_I$=1 (-1) and
 substituting the known analytic solution for the instantons one finds
\beq     {1 \over 16 \pi^2 }   
            G_{\mu\nu} {\tilde G_{\mu\nu}}   =  
       Q_I    {1 \over \pi^2}  { 12 \rho^4 \over
      (\rho^2 + (y-z_I)^2 )^4 }
      \label{ggmu}
\eeq
and
\beq   { 1 \over 16 \pi^2 m_c^2 }  f^{abc} 
        G_{\mu\nu}^a {\tilde G}_{\nu\alpha}^b G_{\alpha\nu}^c 
          = -
              Q_I       { 96 \over \pi^2 }  {1 \over m_c^2 } 
           { \rho^6  \over
             \left( \rho^2 + (y-z_I)^2  \right)^6    } 
\label{charm2}
\eeq  
where $\rho$ is the size and the $z_I$  the position.
The first important observation now is that each instanton/anti-instanton 
contributes to $\Delta \Sigma$ and $\Delta c$ with the {\it opposite} sign. 
%


%
The second is that, by dimension, charm component relative to light quark one
is $\sim <\rho^{-2}>/m_c^2$, where angular brackets mean averaging over
instanton size distribution. Obviously this average favor smaller sizes than
the usual r.m.s. one. 

  Of course, due to the factor of the 
topological
charge $Q_I=\pm 1$ the average of both of those
pseudo scalar operators in the vacuum vanishes.
Non-zero results can only be obtained if it is correlated with another
pseudo scalar observable, such as $\vec\sigma \vec q$, a product of spin 
vector
and the momentum transfer. Another useful way to explain it is to  
note that a spin vector acts
on the vacuum like a dipole operator, thereby polarizing (topological) 
charges and creating non-equal densities of instantons and anti-instantons
along its vector. 

4. Quantitative calculations are based on 
 the three-point correlation functions 
indicated above. Instead of using the axial current
itself, we are considering its divergence \queq{diveq} in \qeq{corrfun}.
Then for x in Euclidean time direction, the spin asymmetry 
induced by  
the pseudoscalar
gluonic operators in the direction n of y can be traced out with 
$\Gamma^{n -}$. The correlator is then $\sim (y\cdot  s)$ which
translates into $\sim (q\cdot s)$ in momentum space. Since we plot
the correlator in coordinate space as a function of y 
we are not facing the 
problem of a vanishing forward matrix
element \cite{ins:jama} of these operators.

For comparison, we use
 $two$ instanton ensembles, 
 the random instanton liquid (RILM) 
and the so called
interacting instanton liquid model (IILM) (for review see \cite{ins:report}).
The former is a simple model based on phenomenological parameters of the
instantons like the average size, while the latter is a sophisticated
model including the 't Hooft effective interaction to all orders. 
The ensembles have
256 instantons in a rectangular box of $(5.7\fm)^2\times(2.8\fm)^2$.
The instanton density is the same 
$n=1\ \fm^{-4}$ for all configurations.

  Correlations functions with the nucleon propagator decay very fast
with x due to
large nucleon mass, which in practice strongly
 restricts the  range of x,y at which
the measurements can be made. Virtual charm pairs are rather well localized
anyway,
and so one may probably use a single ``constituent'' quark\footnote{More
 complete analysis of structure functions of a constituent quark
resulting from those measurements will be presented elsewhere \cite{BS97zz}.}
 instead of three
in the nucleon as a reasonable first approximation.
That is what we did below.
So we do not actually discuss nucleons, but 
some ``generic'' hadrons,  made of
loosely
bound constituent quarks.

Our results for both correlators are shown in Fig.  1 and Fig. 2. The
 geometry of the x and y directions and their ratio we have adopted from
 previous
three-point correlation function studies \cite{Blotz:1997a}: 
their  particular  choice  is not
important for the conclusions.
The dashed lines are guiding analytic calculations based on the
$single$ instanton approximation, 
and serve only as a benchmark for the 
numerical data at
 smaller distances ($x,y\ll 0.5\fm$). 
 
The main finding is that both operators show very similar shape of the
 correlation functions. This is further illustrated in Fig.3, where we plot
 their ratio.
The instanton
ensemble used however matters: for IILM the charm-related operator shows
a  much
 weaker signal. This is related to different  instanton size 
distributions: in RILM all of them have $\rho=1/3 \fm$ while in IILM the
 distribution is
peaked at about $0.4 \fm$.  
Another important reason for the reduction effect in IILM  are positive
(screening-type) correlations between instantons and anti-instantons induced by
quark exchanges.
We use the similarity in shape of both correlators as a sign that 
the form factors for both $\Sigma(q^2)$ and $\Delta c(q^2)$ are the same, 
and the ratio
of the  correlation functions  shown in Fig.3 can be  immediately
translated to
the ratio ${\Delta c\over\Sigma}$.

Finally our results are 
\beq    {\Delta c\over\Sigma} \simeq 
     - 0.20 \pm 0.04 
     \,\,\, (RILM),
\label{ratrilm}   \eeq
and  
\beq {\Delta c\over\Sigma}\simeq  
      - 0.08  \pm 0.006
     \,\,\, (IILM).\label{ratiilm} \eeq  

Several things have to be remarked now.
Using an approximate scheme for the non-zero mode propagator
involved in the correlation function would give the analytic expression 
${\Delta c\over\Sigma}=-{12\over 5N_f} {1\over (m_c\rho)^2}\simeq -0.2$ 
\cite{BS97zz} for the standard values of the RILM. This scheme is 
also shown in Fig.3 and actually very close to the full RILM simulation.  

The positive sign of $\Delta \Sigma$ here agrees with the experimental
data and the negative sign for $\Delta c$ is somehow in agreement with
lattice calculations \cite{Liu:1995kb} of disconnected matrix element of the
axial current itself, which give a negative sign for the three light
quark flavors.
Both $\Delta\Sigma$ and $\Delta c$ signs are opposite however to the 
scenario proposed by Halperin and Zhitnitsky
\cite{Halperin:1997b}, which is mainly
based on an older low energy theorem by K\"uhn and Zakharov
\cite{Kuhn:1990}. 

These numbers (\ref{ratrilm},\ref{ratiilm}) are significantly smaller than the 
crude estimates for
$g_A^{(0)}$ based on Goldberger-Treiman-type  
relations for $\eta'$ exchange used in \cite{shuzhi:1997}.  
Nevertheless the size of $\Delta c$ is in fact surprisingly large. It seems
to be only 3-6 times smaller than that of the strange quark sea. 
Hopefully the next generation
of polarized DIS experiments \cite{Baum:1996,Mallot:1996rd}
with charm-jet tagging will be able to observe
it.

5. Acknowledgments. We thank P.van Baal, the organizer of the NATO school in
 Cambridge, England, in June 1997, 
for invitations: the present work was initiated 
there. One
of us (ES) is much indebted to A.Zhitnitsky 
who has shared his ideas with him and 
for pointing out an error in an earlier version of this paper.
ES and AB acknowledge partial support by US Department of Energy and 
AB thanks A.von Humboldt 
foundation for a past Feodor Lynen fellowship.


\pagebreak 

\newpage 
\hskip 1.9in    
 \begin{center} 
 \epsfig{file=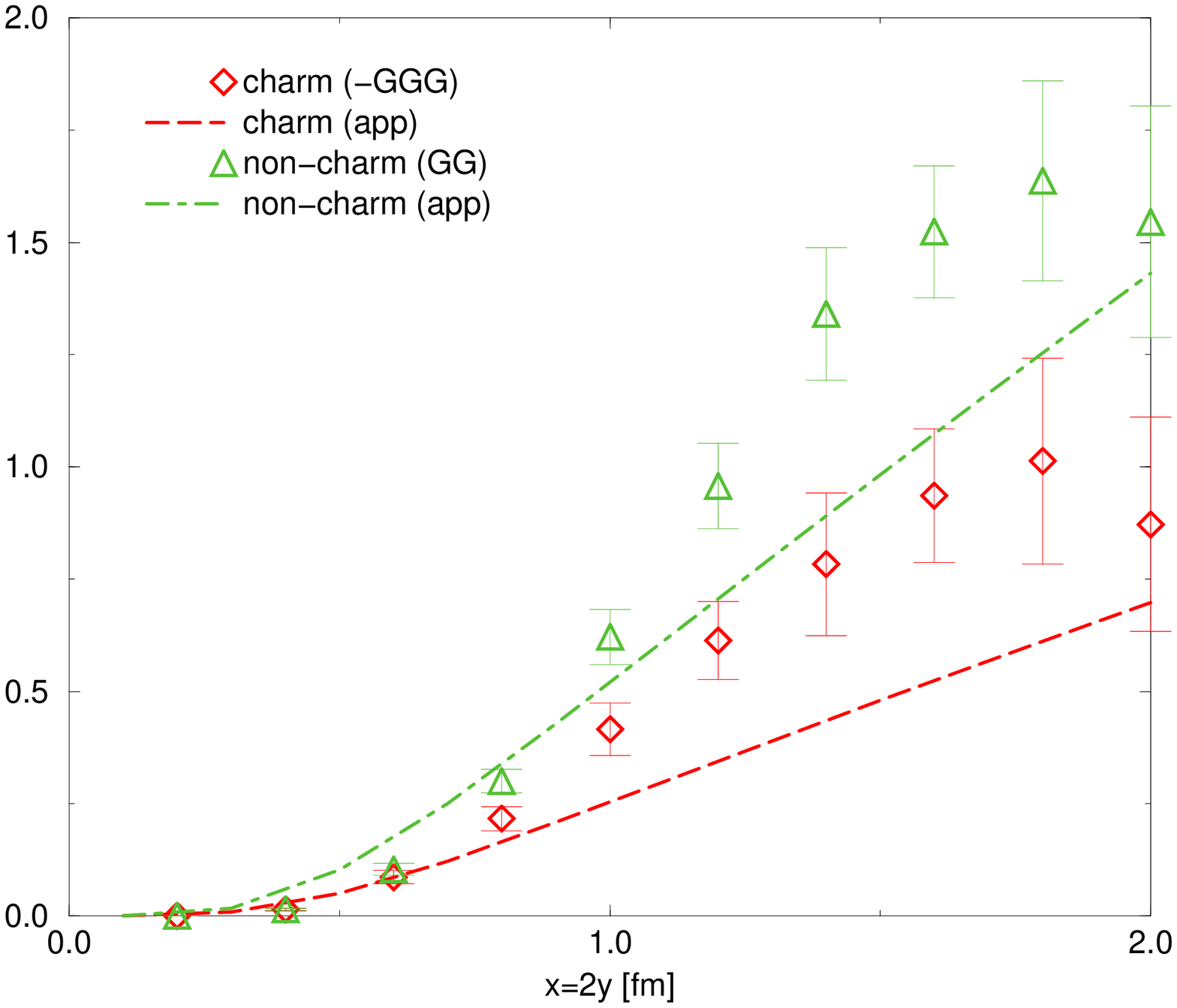,width=18cm,angle=-90}
 \parbox{6in}{ \label{charmeps}
 \phantom{abc}
 \hskip0.5cm 
 \small Fig.~1:\quad The quark three-point correlation function  
for the charm contribution $\simeq f^{abc} G^a(y){\tilde G}^b(y)G^c(y)$  
for the random  
configuration 
(RILM). The approximate (app) formula is also shown. }  
\end{center} 
\newpage 
\hskip 1.9in    
 \begin{center} 
 \epsfig{file=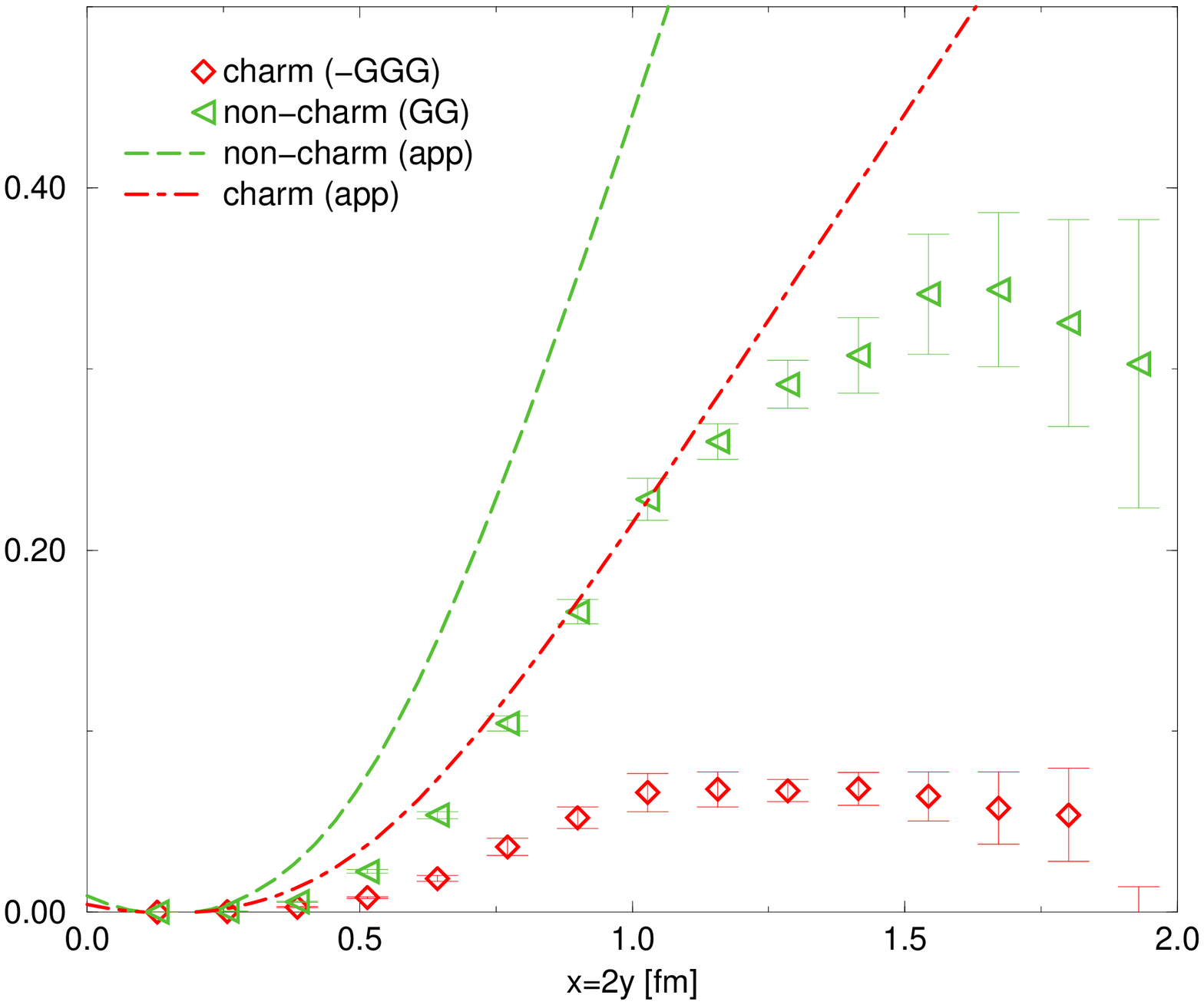,width=18cm,angle=-90}
 \parbox{6in}{ \label{charmieps}
 \phantom{abc}
 \hskip0.5cm 
 \small Fig.~2:\quad The quark three-point correlation function  
for the charm contribution $\simeq f^{abc} G^a(y){\tilde G}^b(y)G^c(y)$  
for the interacting  
configuration 
(IILM).  } 
\end{center} 

\newpage 
\hskip 1.9in    
 \begin{center} 
 \epsfig{file=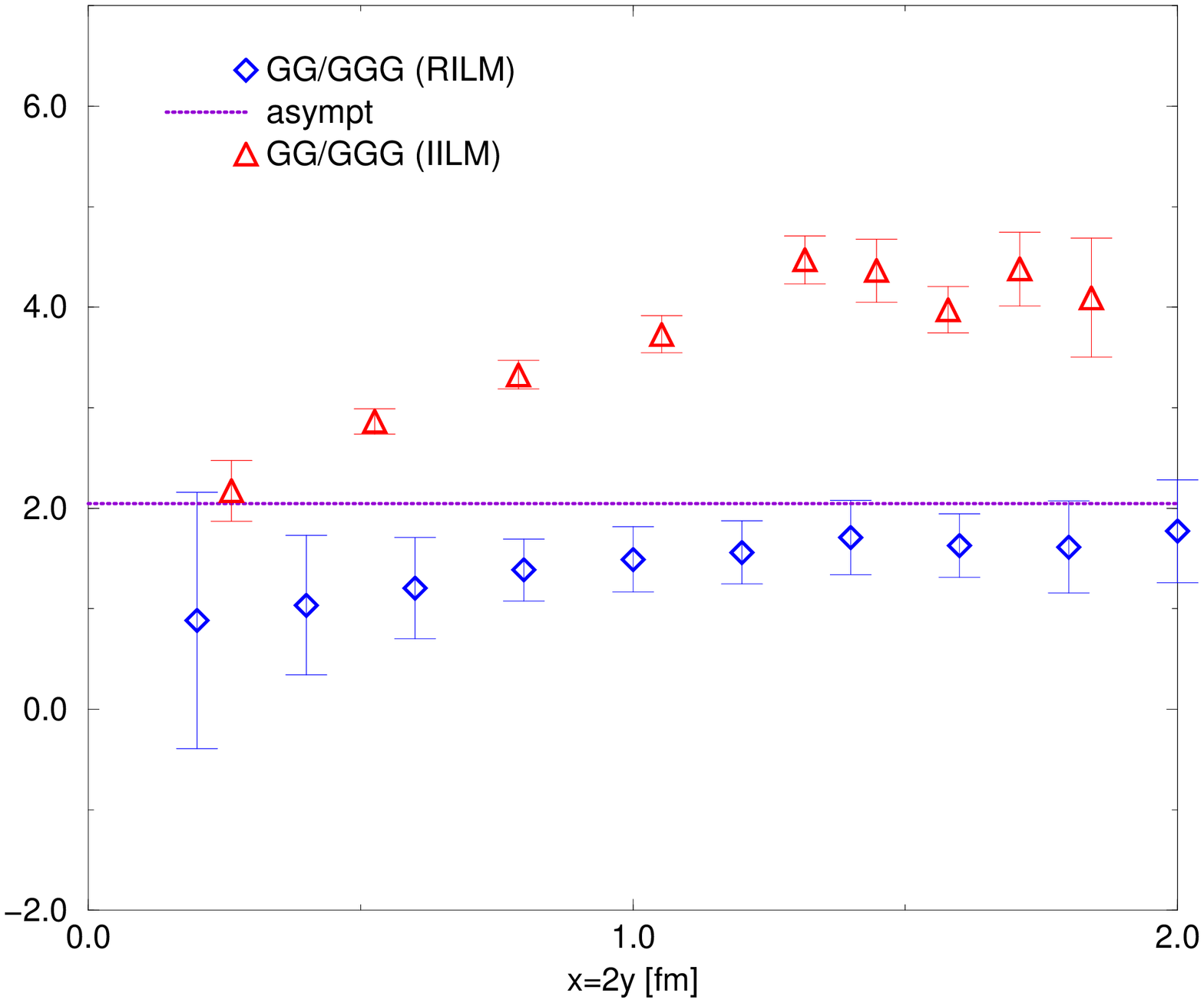,width=18cm,angle=-90}
 \parbox{6in}{ \label{ratioeps}
 \phantom{abc}
 \hskip0.5cm 
 \small Fig.~3:\quad The ratio of the three-point function 
for ${N_f\over 16 \pi^2} G{\tilde G}$ and 
$-{N_f\over 16\pi^2}  f^{abc} G^a(y){\tilde G}^b(y)G^c(y) $ 
for the RILM and IILM configuration. } 
\end{center} 

\end{document}